\documentclass[preprint2]{aastex}

\usepackage{hyperref}

\shorttitle{Rotation Velocity Lags}
\shortauthors{Bizyaev et al.}

\begin{document}

\title{SDSS IV MaNGA - Rotation Velocity Lags in the Extraplanar Ionized Gas from MaNGA Observations of Edge-on Galaxies}

\author{Bizyaev, D. \altaffilmark{1,2,3} , 
Walterbos, R. A. M.\altaffilmark{4} ,
Yoachim, P. \altaffilmark{5},
Riffel, R. A.\altaffilmark{6,7},
Fern\'andez-Trincado, J. G. \altaffilmark{8},
Pan, K. \altaffilmark{1},
Diamond-Stanic, A. M.\altaffilmark{9,10},
Jones, A.\altaffilmark{11},
Thomas, D. \altaffilmark{12},
Cleary, J. \altaffilmark{13},
Brinkmann, J.\altaffilmark{1}
}

\altaffiltext{1}{Apache Point Observatory and New Mexico State University, Sunspot, NM, 88349, USA}
\altaffiltext{2}{Sternberg Astronomical Institute, Moscow State University, Moscow, Russia}
\altaffiltext{3}{Special Astrophysical Observatory of the Russian AS, 369167, Nizhnij Arkhyz, Russia}
\altaffiltext{4}{Department of Astronomy, New Mexico State University, Las Cruces, NM, 88003, USA}
\altaffiltext{5}{Department of Astronomy, University of Washington, Seattle, WA, 98195, USA}
\altaffiltext{6}{Departamento de F\'isica, CCNE, Universidade Federal de Santa Maria,
Av. Roraima, 1000 - 97105-900, Santa Maria, RS, Brazil}
\altaffiltext{7}{Laborat\'orio Interinstitucional de e-Astronomia - LIneA, Rua Gal.
Jos\'e Cristino 77, Rio de Janeiro, RJ - 20921-400, Brazil}
\altaffiltext{8}{Institut Utinam, CNRS UMR 6213, Universit\'e de Franche-Comt\'e, OSU
THETA Franche-Comt\'e-Bourgogne, Observatoire de Besancon, BP 1615, 25010
Besanon Cedex, France}
\altaffiltext{9}{Department of Astronomy, University of Wisconsin-Madison, Madison, WI 53706, USA}
\altaffiltext{10}{Department of Physics and Astronomy, Bates College, Lewiston, ME 04240, USA}
\altaffiltext{11}{Max-Planck Institute for Astrophysics, Karl-Schwarzschild-Str 1, Garching, Germany 85748}
\altaffiltext{12}{Institute of Cosmology and Gravitation, University of Portsmouth, Dennis
Sciama Building, Portsmouth, PO1 3FX, UK}
\altaffiltext{13}{Department of Physics \& Astronomy, Johns Hopkins University,  Bloomberg Center, 3400
N. Charles St., Baltimore, MD 21218, USA}

\begin{abstract}

We present a study of the kinematics of the extraplanar ionized gas
around several dozen galaxies observed by the Mapping of Nearby
Galaxies at the Apache Point Observatory (MaNGA) survey.  We
considered a sample of 67 edge-on galaxies out of more than 1400
extragalactic targets observed by MaNGA, in which we found 25 galaxies
(or 37\%) with regular lagging of the rotation curve at large
distances from the galactic midplane. We model the observed $H\alpha$
emission velocity fields in the galaxies, taking projection effects and
a simple model for the dust extinction into the account. We show that
the vertical lag of the rotation curve is necessary in the modeling,
and estimate the lag amplitude in the galaxies.  We find no 
correlation between the lag and the star formation rate in the
galaxies. At the same time, we report a correlation between the lag
and the galactic stellar mass, central stellar velocity dispersion,
and axial ratio of the light distribution. These correlations suggest
a possible higher ratio of infalling-to-local gas in early-type disk
galaxies or a connection between lags and the possible presence of hot
gaseous halos, which may be more prevalent in more massive
galaxies. These results again demonstrate that observations of extraplanar gas
can serve as a potential  probe for accretion of gas.

\end{abstract}


\section{Introduction}

Warm, ionized gas at high altitude from the midplane has been detected
in the Milky Way
\citep{hoyle63,reynolds71,reynolds73,reynolds99,shull09}. It is also
observed in other galaxies
\citep{dettmar90,rand90,rand97,hoopes99,rand00,rossa04} at the
distances up to several kpc \citep{rossa03,wu14} from galactic
midplane.  It is one of the tracers indicating significant amounts of
extra planar gas.
This material is envisioned as an important component of
galactic evolution that probes disk-halo interaction \citep{putman12}.  The material
may help support and regulate star formation in galaxies via
replenishing gas close to the midplane.  

The origin of the
extrafplanar gas is still unclear. It may have contributions from two
sources: cosmological accretion from the intergalactic medium
reservoir \citep{binney05}, or gas associated with galactic fountain
flows in galaxies that are produced by star formation in the disk
\citep{shapiro76,bregman80,norman89,marinacci11,fraternali13}. 
Both of these effects may be at
work together \citep{haffner09,benjamin12}.  
Since the gas flow motion is either parallel to the sky plane
for edge-on galaxies, or obscured by galactic midplane gas, it is difficult to
establish from observations if          the gas 
is on the way out or in, caused by fountain flows, or represents some
quasi-stable phase of gas energized by continuous activity in the
disk. 
The quantity and distribution of the material suggests the latter, 
but it
does not imply that a connection to flows is not present. Different
classes of objects (e.g. AGNs, young stars, old evolved hot stars, or 
dynamical features) can be responsible for the extra planar gas
ionization and motion.
Circulation of the gas between the galaxy and the intergalactic medium
may be driven by star formation in
the disk.

A deviation of the gas component's rotation from cylindrical at large
distances form the midplane, i.e. the rotation velocity lag, or
briefly ``lag'', 
has been noticed in several nearby galaxies.  
It has been measured in the atomic gas 
(see e.g. \citet{swaters97,matthews03,zschaechner11,gentile13,kamphius13,zschaechner15a,zschaechner15b}), 
and the ionized gas (\citep{fraternali04,heald06a,heald06b, heald07,kamphius07,kamphius11,rosado13,wu14,boettcher16}).
The number
of galaxies with measured rotation velocity lags is still small, and
the lack of data hinders statistical studies. It has been known for a while that a
strong correlation exists between star formation per unit disk area
and the presence of ionized extra planar gas \citep{dettmar04,ho16},
but fewer galaxies have measurements of kinematic lags.  An inverse
correlation between the lag amplitude and the star formation activity
in galaxies was suggested by \citet{heald07} but not confirmed by
\citet{zschaechner11}.  In this paper we define lag as a decrease in rotational 
velocity with height.

Studying correlations between the lag and general galactic parameters
can be a powerful tool for understanding the nature of the
extraplanar gas. 
The most recent results \citep{zschaechner15a}
point at a radially decreasing lag amplitude in galaxies, which the
authors interpret as likely pointing at an internal origin of the
lags.  
Lag measurements can also provide potential evidence
for accretion of gas onto galactic disks (e.g. \citet{fraternalli06,
  fraternali08}. Large Integral Field Unit surveys being conducted in
recent years (e.g.   
CALIFA\footnote{Calar Alto Legacy Integral Field spectroscopy Area survey}, \citet{califa}, 
MaNGA\footnote{Mapping Nearby Galaxies at Apache Point Observatory}, \citet{bundy15}, and
SAMI\footnote{Sydney-Anglo-Australian Observatory Multi-object Integral field survey}, \citet{sami}) 
provide excellent data from panoramic spectroscopy
suitable for studying the extraplanar gas component in various types
of galaxies. In this paper we employ a growing IFU data set from MaNGA
survey for studying the kinematics of ionized gas around a substantial
sample of edge-on disk galaxies.

\section{MaNGA observations}

MaNGA is a massive integral field unit (IFU) spectroscopic survey
\citep{bundy15,drory15} that is part of the fourth phase of 
the SDSS\footnote{Sloan Digital Sky Survey, http://sdss.org}
survey \citep{blanton16}. MaNGA plans to deliver R$\sim$2000
spectroscopic maps for about ten thousand nearby galaxies by
2020. MaNGA uses the Sloan 2.5 m telescope at Apache Point Obervatory
\citep{gunn06} and spectrographs that cover the
3600-10300 \AA~ spectral range \citep{smee13}. The survey's target selection
\citep{law15} and observing strategy provide a kiloparsec-scale
spatial resolution maps of the stellar and ionized gas kinematics in
the galaxies: the resulting FWHM $\sim$ 2.5 arcsec \citep{law16}
 in the parameter maps corresponds to 1.5 kpc at the median
redshift of the survey of 0.03.  The observing strategy provides
contiguous coverage of targets, without leaving unobserved gaps. The
spectra are flux calibrated to the precision of a few percent
\citep{yan16}.  Current data release MaNGA Product Launch - 4 (MPL-4,
Westfall et al., in prep., \citet{law16}) provides spectral
data cubes and various parameter maps derived from them for more than
1400 unique galaxies (including ancillary targets).

\subsection{The Sample of Edge-on Galaxies in MaNGA}

Since all MaNGA galaxies were selected based on SDSS imaging products
\citep{law15}, we were able to download and inspect the
color-composite images of all galaxies from SDSS SAS\footnote{Science Archive Server} 
archive \citep{sdssrd12,blanton11}.  We identified edge-on galaxies from
visual inspection similar to how it was done in \citet{EGIS}: objects
were classified as edge-on systems if they had no traces of spiral
arms that could be seen, and also if the dust lane (if present) was
projected on or very close to the galaxy's disk midplane. We also
restricted our search to objects not showing obvious signs of
interaction.  As a result, we identified 67 edge-on galaxies, without
any additional selection by morphological type, size, etc. MaNGA IFUs
range in size from 12 arcsec (19-fibers) to 32 arcsec (127-fibers), see
the IFU description in \citet{drory15}.  Out of the 67 selected
edge-on galaxies, 49\% were observed with the largest, 127-fiber IFUs,
the other 9, 24, 13, and 5\% were observed with 91-, 61-, 37-, and
19-fiber IFUs, respectively.
 
\subsection{Selecting Edge-on Galaxies with Measurable Rotation Velocity Lag}

The MPL-4 release provides gas radial velocities estimated for
different emission lines on a spatial grid of 0.5 arcsec spaxel
size. We utilize H$\alpha$ fluxes with uncertainties to estimate the
signal-to-noise ratio (SNR), and H$\alpha$ radial velocities with
uncertainties for drawing maps of the gas kinematics.  In
the majority of galaxies the H$\alpha$ fluxes are detected at
higher than 1 kpc altitudes from their galactic midplane. Then we
considered the vertical profiles of the radial velocities at different
radii from the center, typically in 1-2 kpc wide bins along the major
axis.  Only the spaxels with good data quality flags and with SNR$>$3
were taken into account.  We find that 42 objects in the sample show a rather
irregular shape of the vertical velocity profiles, whereas 25 galaxies
show a regular decreasing velocity amplitude, i.e. lag, with vertical
distance from the midplane, at all radial distances from the center 
within the limits of our angular resolution of 2.5 arcsec.
We refer to this group of galaxies as subsample in the text below, to distinguish it
from the main initial sample of edge-on galaxies.

\begin{figure}
\epsscale{1.00}
\plotone{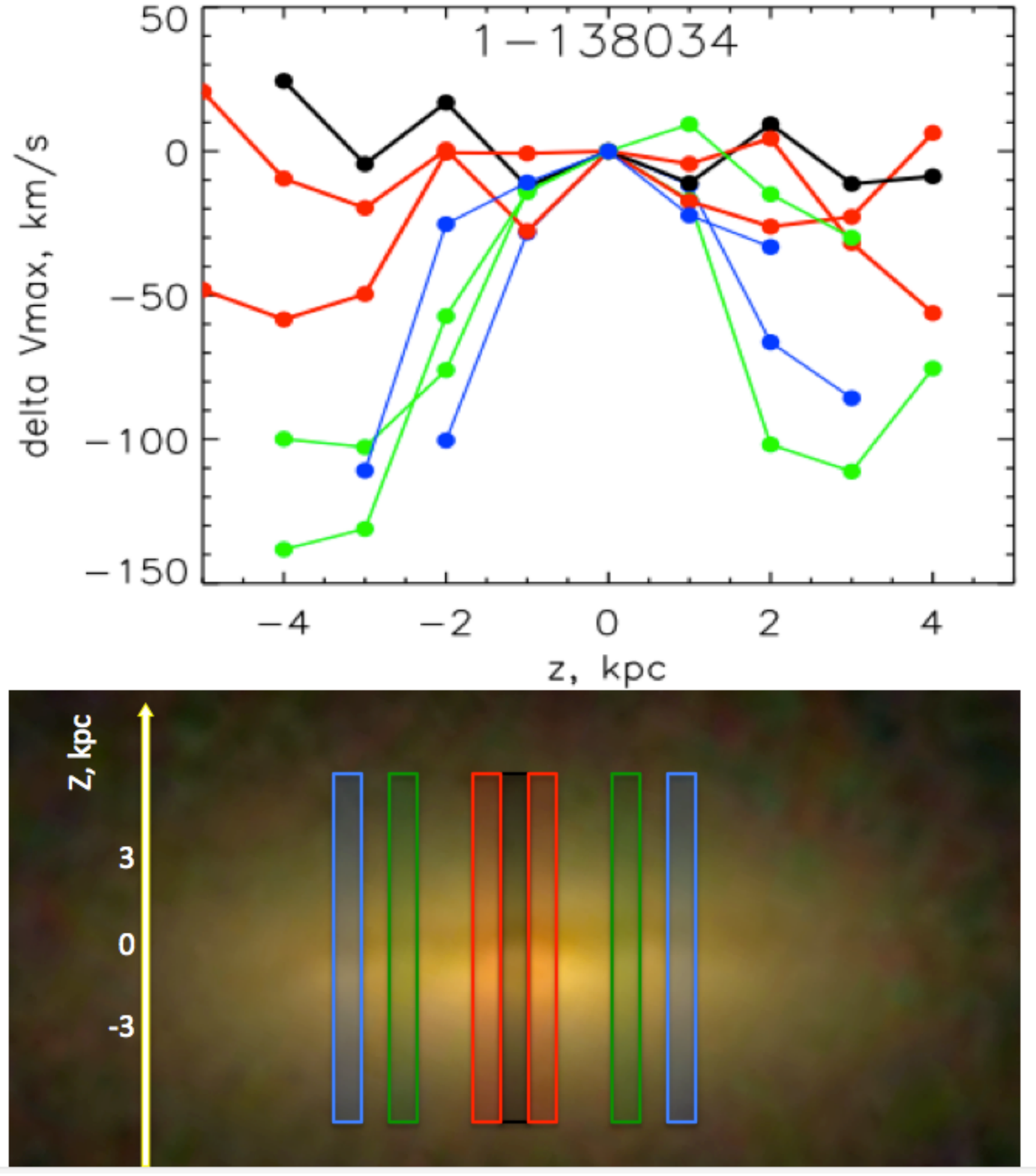}
\plotone{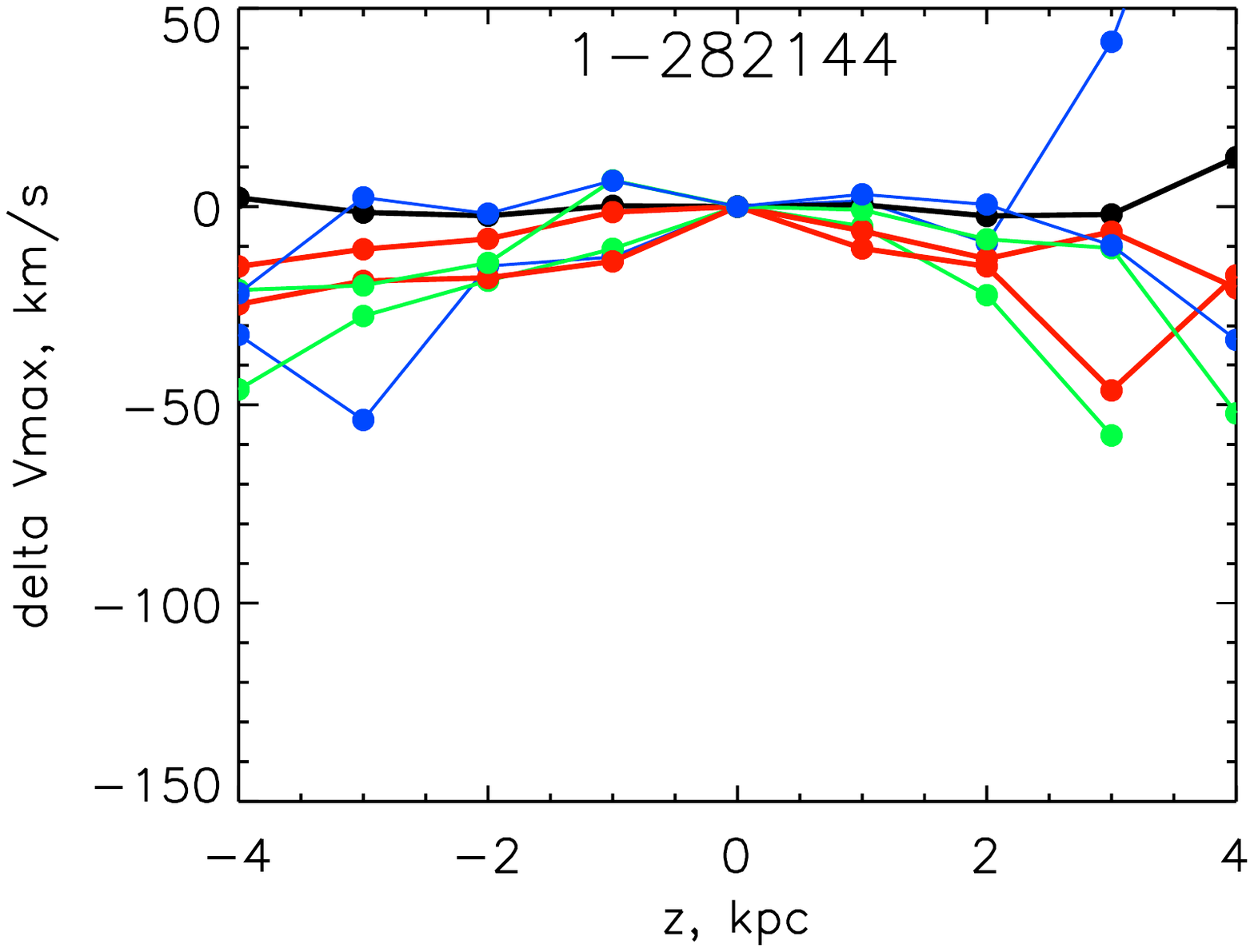} 
\caption{Vertical profiles of radial velocity in two galaxies from our sample.
  Top panel: the galaxy 1-138034 is shown as a typical example.  
  All velocities are shown with respect to the
  galactic midplane velocity.  
  The black curve designates
  the velocity vertical profile that goes through the center (X=0).
  The red curves are for X=0.8 kpc, the green ones correspond to X=3
  kpc, and the blue ones designate X=4 kpc.  Here X designates the
  projected distance to the center along the major axis on the edge-on
  galaxy.  
  Middle panel: regions used for making the radial velocity
  profiles in the galaxy 1-138034 designated with the same colors as in the top panel. The
  background image of the galaxy is taken from SDSS SAS server (see
  http://sdss.org).
  Bottom panel: same vertical profiles of radial velocity as in the top panel shown
  for the galaxy 1-282144, in which the lag is seen less clearly.
\label{fig1}}
\end{figure}

Figure~\ref{fig1} shows two examples of galaxies with vertical velocity
profiles that suggest that the rotation curve has lower amplitude
above the midplane than in the midplane, i.e. 
that we observe a lag. The top panel in Figure~\ref{fig1} is 
a typical galaxy in our sample, while the bottom panel
in Figure~\ref{fig1} shows a galaxy with a less clear lag. 

We cannot calculate
the lag amplitude directly from Figure~\ref{fig1} because of
projection effects and possible dust extinction. It is worth noting
however, that these effects would not decrease the rotation curve
amplitude systematically with altitude above the galactic midplane in
the case of pure cylindrical rotation of gas, i.e. in the case of no
lag. Thus, we assume that the regular decrease in the gas rotation
velocity with vertical distance to the midplane is an indication of a
lag.  We selected 25 galaxies with regular decreasing velocity
amplitude with vertical distance to the midplane for the further
modeling. 

\begin{figure}
\epsscale{2.4}
\plotone{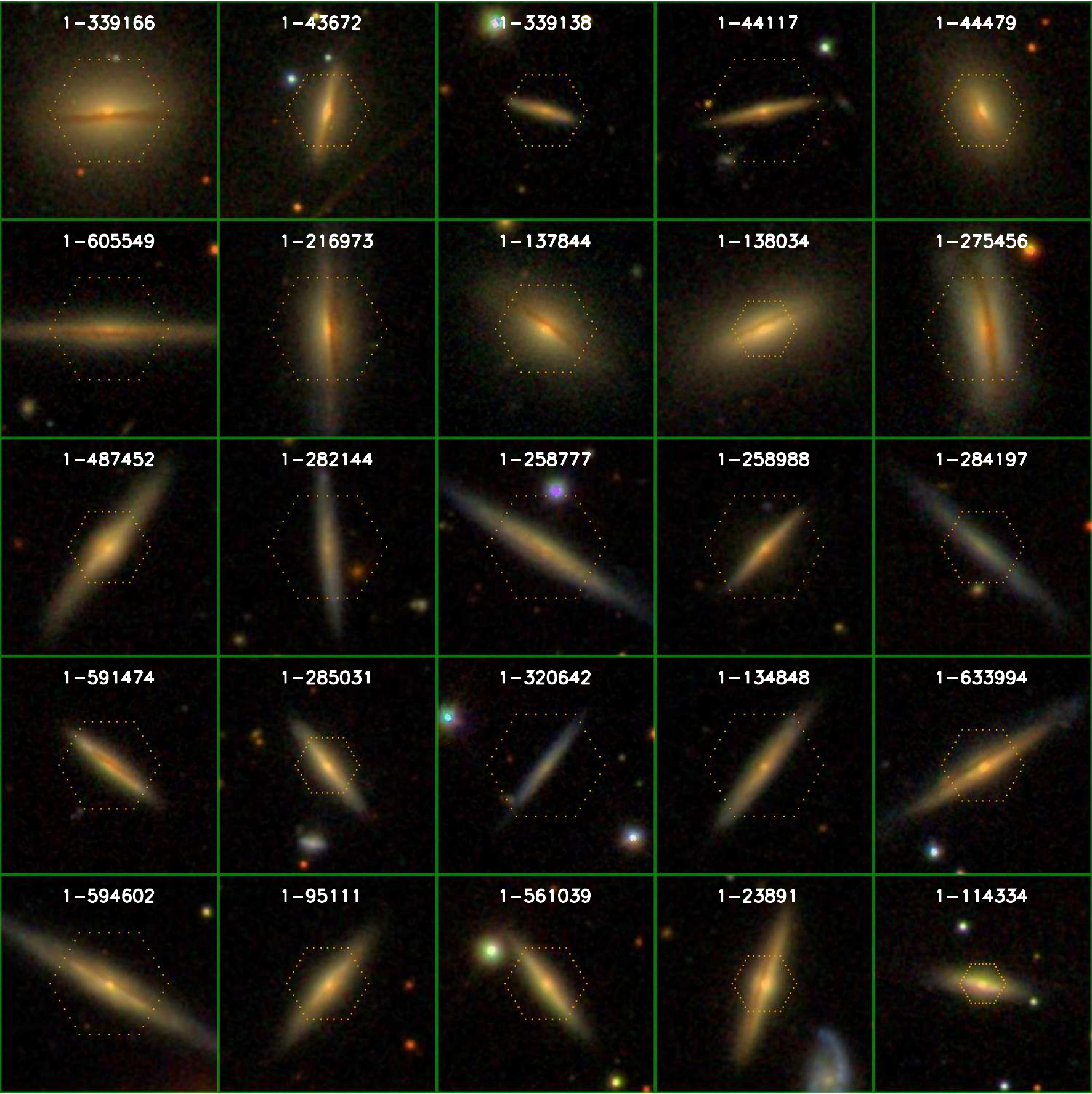}
\caption{
Images of selected 25 galaxies with regular lag (see \S2.2).
The images are taken from the SDSS SAS server (http://sdss.org).
All objects are shown on the same scale: each individual image has 
1 arcmin x 1 arcmin size. The order of the objects is the same as in
Table~\ref{tab1}: from the top row left to right corresponds
to lines 1,2,...5 in Table~\ref{tab1}. The 2nd top row in the figure corresponds
to lines 6,7,...10 in Table~\ref{tab1}, etc. Table~\ref{tab1} is introduced
in \S{3}, see below. The yellow dotted contours show the allocation of MaNGA IFUs
on the galaxies. 
 \label{fig1b}}
\end{figure}

Images of selected galaxies are shown in Figure~\ref{fig1b}.
It is interesting to note that the IFU size distribution for this
subsample well matches that of the overall sample of 67 MaNGA edge-on
galaxies: 54, 13, 21, 8, \& 4\% for 127-, 91-, 61-, 37-, and 19-fiber
IFUs, respectively. We will come back to the 42 galaxies that do not
show a systematic lag in the discussion in Section 4.

The panels in Figure~\ref{fig1b} are 1 arcmin on the side. The largest IFU
bundle used for 13 of the 25 galaxies is about half that size. Thus,
none of the galaxies are fully covered in the radial extent visible in
the figure. This fact implies that we probe the lag mostly over the inner disk
and cannot address with these data if there are radial variations in
the lag. 
At the same time, the data probe well outside the central region 
and the galactic midplane.

\clearpage
\clearpage

A single fiber corresponds to 1.5 kpc at the median survey redshift. The
smallest fiber bundle for this sample (19 fibers) still has 5 fibers
across each galaxy. The largest bundle used for more than half the
subsample has from 11 to 13 fibers across the major axis of each galaxy,
depending on fiber bundle orientation. 
Coverage decreases away from
the center of the galaxy due to the hexagonal shapes of the fiber
bundles so the radial extent over which lags may be probed is somewhat
smaller than the major axis extent \citep{drory15}.

\section{Modeling Ionized Gas Velocity Fields}

Below we assume that the ionized gas that emits $H\alpha$ follows a
bi-exponential 3-dimensional distribution in the galaxy, and introduce its
rotation curve as an analytical function.  We construct model maps of
the two-dimensional distribution of $H\alpha$ radial velocities by
integrating the volume density weighted $H\alpha$ emission profiles
along the line of sight. Then we compare the resulting $H\alpha$
velocity fields with observations and obtain the model parameters via
chi-square optimization.

The model rotation curve in the midplane is approximated by three
parameters: 
radius $r_0$ to which the velocity rises linearly,
the rotation velocity at $r_0$, $V_{max}$, and the gradient of the
rotation curve beyond the $r_0$, $dv/dr$. The latter allows for the
possibility of a slightly rising or decreasing rotation curve. Hence,
$V(r) = V_{max} \cdot (r/r_0)$, if $r ~<~ r_0$, and $V(r) = V_{max} +
dv/dr \cdot (r - r_0)$, if $r ~\ge~ r_0$, where r is the distance to
the galactic center.


The lag of the rotation velocity is
introduced as a linear vertical gradient $dv/dz$: $V(r,z) = V(r)  -
|z| \cdot dv/dz$, where z is the vertical distance in kpc. The
rotation curve and the lag are assumed independent of the polar angle
in the disk plane, $\phi$.  Since the observed rotation curve is a
result of projection of $V(r,\phi,z)$ on the sky plane, we have to
consider the line-of-sight integration of $V(l,z)$ weighted by the
H$\alpha$ luminosity volume density $f_{H\alpha}$.  Here $l(r, \phi)$
designates the distance along a particular line of sight.

We assume that the distribution of the H$\alpha$ luminosity density is
a double exponential: $f_{H\alpha}(r,z) = f_0 \, exp(-r/h) \,
exp(-|z|/z_0)$, where the scale length $h$, scale height $z_0$, and $f_0$  are
estimated from the $H\alpha$ maps of the galaxies using the structural
parameters pipeline used in \citet{BM02,BM09,EGIS}.  The $H\alpha$ maps are
obtained from the fluxes provided by MPL-4.

We also introduce a dust disk embeded in the galactic disk (co-planar
with the ionized gas and stellar disks). The dust extinction
coefficient $\kappa (r,z) = \kappa_0 \, exp(-r/h_d) \, exp(-|z|/z_d)$
is the volume extinction coefficient at the $H\alpha$ wavelength in a
certain point ($r,z$) in the disk, $h_d$ and $z_d$ are the radial and
vertical disk scales, and $\kappa_0$ is the value of the extinction
coefficient in the center. We neglect dust scattering in the sense
that we treat $\kappa$ as a volume absorption coefficient and ignore
the possibility that scattered light might contribute along a line
of sight.

We assume that each point (r,$\phi$,z) in the model disk produces an
emission line profile with the central wavelength that corresponds to
its velocity V(r,z) and the peak height proportional to the luminosity
volume density. We assume the same velocity dispersion sigma of
$\sigma_w =$10 km s$^{-1}$ for all local Gaussian profiles at each
(r,$\phi$,z).  For the radial velocity profile in each point in the
disk $v_p(w, r, \phi,z)$, we assume that $v_p(w,r,\phi,z) = V(r,z) \,
exp(-w^2/2\sigma_w)$.  Here $w$ is the offset of the radial velocity
from the peak velocity V(r,z), and $w$=0 at the peak.  If X and Y
are the radial (along the major axis) and vertical (along the minor
one) coordinates in the sky plane, the integrated velocity profile
$V_p$ can be written
\begin{equation}
V_p(w, X, Y) = \int \limits_{-L}^{L} v_p(w, r(l),  z) \, cos(\phi(l))\, f_{H\alpha}(r(l),z) \, e^{-\tau(r(l),z)} \, dl \,\, ,
\end{equation}
where

\begin{equation}
\tau(r(l),z) = \int \limits_{l}^{L}  \kappa (r(l),z) \, dl   \, ,
\end{equation}
and $cos(\phi(l))$ accounts for the projection of the velocity
vector on the line of sight, and the integration along the line of
sight is towards our position starting from the far side of the disk.
The vertical axis $z$ is aligned with the sky plane axis Y, so $Y =
z$.  The limits of integration in $L$ correspond to $r = 4 \, h$,
by analogy with stellar disks that truncate at their
four radial scales, on average.

After the integration along the line of sight we obtain a resulting
emission line profile $V_p(w)$ of the radial velocity spaced at a set
of spaxels (X,Y). The emission line profile is convolved with a 
Gaussian kernel with FWHM of the spectral resolution. 
To obtain the final "observable" distribution of the
radial velocities in the sky plane, we convolve the $V(X,Y)$ with a
two-dimensional Gaussian point spread function (PSF) whose FWHM
corresponds to the effective FWHM estimated by the data reduction
pipeline for each data cube \citep{law16}. Then we estimate
the corresponding radial velocity $V(X,Y)$ by fitting the radial
velocity profile around the peak velocity value with a Gaussian
function.

We estimate the free model parameters via chi-square minimization
using the downhill simplex "amoeba" optimization algorithm
\citep{amoeba}.
The observing uncertainty of the radial velocity was corrected for the
minimum gas velocity dispersion by adding the expected gas velocity
r.m.s., 10/$\sqrt{3}$ km s$^{-1}$ in quadrature.  We also applied spaxel
quality flags provided by the pipeline: only spaxels with good data
reduction flags and with SNR$>$3 were included into the minimization.
The free parameters of the model are $V_{max}$, $r_0$, $dv/dr$,
$dv/dz$, $\kappa_0$, and also the central radial velocity $V_0$. 
We don't expect any degeneration between the parameters,
so we did not apply any observational constraints to them except 
a natural requirement of the positivity of  $V_{max}$, $r_0$, and $\kappa_0$.
We fixed the dust disk structural parameters as $z_d = z_0$ and $h_d =
h$, which is equivalent to the assumption of uniform mixing of the
ionized gas and dust. For comparison, we ran the same modeling for the
cases of uniformly distributed ionized gas volume density along the
$r$ and $z$. Also we considered shorter vertical and longer radial
scales of the dust $z_d = 0.5 \, z_0$ and $h_d = 1.5 \, h$, and also
$h_d = 2 \, h$ and $h_d = h$
\citep[following][]{xilouris99,kylafis01,matthews99,yoachim06,bianchi07}. In
all cases considered the resulting model parameters were within 15\%
of the original model, which suggests low sensitivity of the output to
the input assumptions on the distribution of ionized gas and dust.
The lag values derived from the model fits are shown in Table~\ref{tab1}.

\begin{figure}
\epsscale{0.90}
\plotone{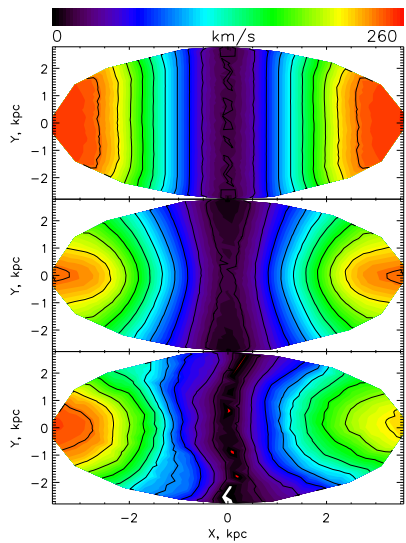}
\caption{Absolute values of velocity fields of our typical galaxy 1-138034: 
model field in the case of fitting with no vertical lag (top), 
model field with lag (middle), 
and the observed velocity field (bottom). 
\label{fig1a}}
\end{figure}

Figure~\ref{fig1a} shows the best-fit model velocity field for the
same galaxy shown in Figure~\ref{fig1} for the case of models with no
lag and with lag (top and middle panels respectively). 
The model simultaneously fits lags in the "left" and "right" parts of the galaxy
(with respect to the minor axis), 
so the absolute values of the radial velocity are shown for clarity of presentation. The
observed velocity field strongly favors the case with the vertical
rotation curve lag.

 
\subsection{Robustness of the Modeling}

\begin{figure}
\epsscale{.70}
\plotone{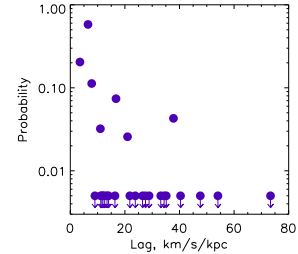}
\caption{Probabilities of the null hypothesis that models with and
  without a rotation velocity lag would statistically result in the
  same chi-square values (i.e. that chi-square does not decrease
  enough to justify the additional parameter to model a lag).  A few
  models with small lags show a high probability for no
  difference, while the bulk of the models gain from using the lag as
  an additional free parameter. The smallest values of the probabilities
  are shown as upper limits at 0.5\% for clarity in the figure.
\label{fig2}}
\end{figure}

To estimate the robustness of our results, we ran the same models
with zero, fixed lag term, $dv/dz$=0.  Then we performed F-tests using
the same sample of galaxies modeled with and without lags.
Figure~\ref{fig2} shows the probability that chi-square values are
significantly similar between the lag and no-lag models.  Note that
most of the probability values are very low, so we truncated the
values in Figure~\ref{fig2} at 0.5\% for better presentation.  As
expected, for several galaxies in the subsample with small lags (10 km s$^{-1}$ kpc$^{-1}$
and less) the chi-square do not change significantly, whereas
models of the galaxies with large lags fit much better when 
a rotation velocity vertical gradient is introduced.

We also constructed a set of synthetic galaxies with typical parameters
$h$, $z_0$, $h_d$, $z_d$, $\kappa_0$, $V_{max}$, $dv/dr$, and
$dv/dz$. For each case the distribution of radial velocity projected
on the sky plane and convolved with observed the PSF was derived. We
added normally distributed uncertainties to the radial velocities in
each spaxel. We also added typical noise to the fluxes in each
spaxel. 
For the latter we used the MaNGA observations completion criterion 
from the science requirement document
\citep{bundy15,law15}, which states that SNR=5 has to be 
obtained at 1.5 $R_e$ (effective radii) for all program galaxies. 
We considered a set of noise levels scaled by SNR = 1, 2, 3, 5, and 10 at
1.5$R_e$. For each set of galactic parameters and noise levels, we
made 30 synthetic galaxies and estimated their parameters with the
pipeline described above.

\begin{figure}
\epsscale{.90}
\plotone{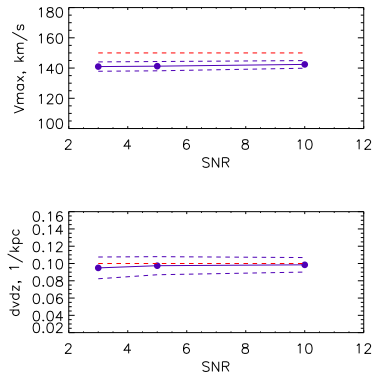}
\caption{The input (red dashed line) and the output model parameters
  (blue curves) estimated for the set of synthetic galactic images
  after inserting noise in them, see text.  The solid curves with
  bullets designate the mean parameters, and the dashed blue curves
  show 1-sigma standard deviation of the parameters in the set of the
  models with the same SNR. The ranges of the vertical axes 
  correspond to the typical values of the parameters in our modeling.
\label{fig3}}
\end{figure}

We obtain that the lag gradient $dv/dz$ is one of
the most reliable parameters estimated by the minimization.  Its
typical uncertainty is 10-15\%. When we convert this uncertainty to the actual lag
in velocity $dV/dz = dv/dz \cdot V_{max}$, the rotation curve
amplitude uncertainty increases the uncertainty on the lag to
20\%. The other reasonably estimated model parameters are the radial
velocity gradient $dv/dr$ and $V_{max}$. The value $r_0$ has
uncertainty of 15\%, but in some cases it is shifted
systematically against the true initial value in different directions
for different model parameters, which increases the uncertainty of this
parameter up to 25\%.
The least reliable parameter is the dust absorption
coefficient $\kappa_0$, which is often underestimated in the modeling,
and its uncertainty sometimes reaches 100\%.  Nevertheless, it does
not affect the inferred rotation curve lag because the high altitude
regions in the galaxies, which are good tracers of the lag, are less
affected by the dust extinction.  Figure~\ref{fig3} shows a result of
the modeling for a typical set of the model parameters.  
We conclude that the uncertainty of our rotation velocity lag 
estimation is 20\%. We are also
inclined to add a low threshold of 6 km s$^{-1}$ kpc$^{-1}$ to this uncertainty based on
Figure~\ref{fig2}, which suggests that given the uncertainty and
resolution of the data, the modeling procedure does not seem sensitive
to variations of the lag below 6 km s$^{-1}$ kpc$^{-1}$.

\subsection{Inclination Effects}

Since our modeling assumes 90$^o$ inclination of galactic disks to the line of 
sight, any deviation from the perfect edge-on inclination may affect parameters
estimated in any studies of edge-on galaxies. \citet{degrijs97} showed that
the photometric vertical scale height increases by 20\% if the inclination $i$
deviates from the 90$^o$ by 5$^o$. We use the visibility of the dust lane 
in SDSS images of all selected galaxies with lags and simple technique
described in \citet{bk04,mosenkov15} to estimate the inclination. 
We do not see any dependence between the lag and the inclination. 
We find that only three galaxies have $i < $85$^o$: 1-561039, 1-339138, and 
1-114334. The latter has $i \approx$82$^o$, and we excluded it from 
the further analysis. The galaxies 1-561039 and 1-339138
have $i$ between 84 and 85$^o$. Although they do not change trends in figures
with the lag shown below in this paper, we excluded them from the figures.

In order to estimate how large the projection due to non-edge-on orientation
is in the galaxies, 
we compare the "projection size" in the galaxies (explained below in this paragraph) 
with the spatial resolution 
of MaNGA data cubes. The latter is assumed to be FWHM/2.35, which corresponds to 
the spatial "sigma" resolution in the MaNGA images,  see \S2). 
The "projection size" is calculated as the diameter of galactic 
disk in projection on the sky plane due to the inclination. 
We find that the overlap between the near and
far sides of galactic disks is less than the spatial resolution in our data for all
objects. Thus we conclude that the inclination effects can be ignored
for our subsample. In general, the inclination affects are important and have to 
be considered when subsamples of edge-on galaxies are formed from
the future MaNGA sample.

\section{Results and Discussion}

We started with a sample of 67 galaxies and found that 25 show
systematic patterns in the velocity field suggesting a lag is present.
Note that we are not claiming that the other 42 galaxies do not show any
lag, only that upon inspection of the velocities it appears less
regular and therefore less suitable to the modeling approach we have
adopted in this paper. 

Previous work in measuring lags for ionized or neutral atomic gas have
involved smaller samples of galaxies. Ours is the first study of a
moderately large sample of observed with one particular IFU setup and
analyzed the same way. The large sample offers opportunities for investigating
trends in lags with integral galactic parameters described below.

The lags are expected as
result of galactic fountain flows, but ballistic models of such flows
indicate much smaller lags than observed 
\citep{collins02,fraternalli06}. An inverse
correlation between the lag amplitude of ionized gas and the star formation activity
in galaxies was suggested by \citet{heald07}.
The inverse correlation was not seen in the neutral gas phase
\citep[see][and references therein]{zschaechner15a}.
Surprisingly, no clear trends have been found
between lag and other galaxy properties so far. 
In particular, lags have not been seen to correlate
with star formation rate, strength of gravity, or
environment. 
The most recent results \citep{zschaechner15a,zschaechner15b} 
point at a radially decreasing lag amplitude in the neutral gas in
galaxies, which the authors interpret as likely pointing at an
internal origin for the lags. Our radial coverage is limited so we
cannot yet address radial variations, but here we consider correlations
with star formation rates and global galaxy properties for our subsample
of galaxies with lag.

To characterize the star formation rate and star formation rate
surface density, we considered the integrated WISE band 4 (22
$\micron$) luminosities for the two subsamples. This WISE band has been shown to
accurately track the SFR \citep[e.g.][]{relano07,zhu08,jarrett13} in
HII regions and galaxies.  The H$\alpha$ luminosities would
underestimate the total SFR due to extinction and lack of coverage of the
outer disk in the IFU spectra. We discuss both effects below for the subsample
of 25. The two subsamples showed similar distributions both in total SFR
and SFR/area. 

We also considered the distribution of axial ratios
(b/a) provided in the NSA for the
galaxies (here we use the axis ratio b/a from Stokes parameters at 50\%
light radius). Interestingly, all galaxies while clearly edge-on or close to
edge-on, show a full range in b/a from 0.3 to 0.9. There is a
difference in the histogram of b/a: the subsample of 42 shows a
peak around b/a of 0.4, which is not seen in the subsample of
25. This difference suggests that the fraction of galaxies with less prominent
bulge/spheroidal components is larger in the sample of 42 
galaxies without regular lags. 

\begin{figure}
\epsscale{.90}
\plotone{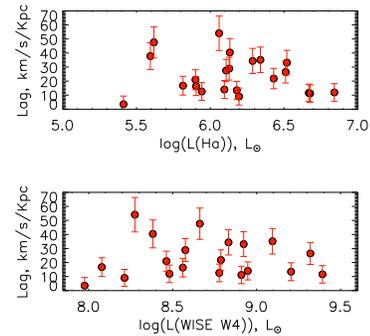} 
\caption{Top: the total $H\alpha$ luminosity in the galaxies and the
  rotation curve lag.  The luminosity is not corrected for the
  internal extinction.  Bottom: Lag versus the luminosity of the
  galaxies in the WISE W4 band. The W4 luminosity should correlate well with the
  star formation rate in the galaxies.
\label{fig4}}
\end{figure}

We will now analyze trends for the subsample of galaxies where we
derived a robust lag value. Figure~\ref{fig4} shows the relationship
between the total $H\alpha$ luminosity and the lag. The luminosity is
estimated by the integration of the total $H\alpha$ flux in the
galaxies from MaNGA data.  As mentioned above, this estimate may be incomplete
due to IFU coverage and it will certainly suffer from extinction. We
see no clear correlation
(with negative correlation coefficient $cc$ = -0.17).  The lack of correlation was
also apparent when we replaced the $H\alpha$ luminosity with the mean
surface density of the the $H\alpha$ luminosity ($H\alpha$ luminosity
divided by the disk area, where the disk area was derived using the
scale length of the $H\alpha$ disk; figure not included).

A better indicator of global star formation in the case of edge-on
galaxies is the infrared (IR) luminosity and in particular the band 4
WISE luminosity, as discussed above. We calculated the IR luminosity
in the WISE W4 band (at 22$\mu$), which is very likely optically
thin. The WISE fluxes for the galaxies are taken from the ALLWISE
catalog \citep{allwise}. We find that the H$\alpha$ and W4
luminosities are linearly correlated, although with significant
scatter. The average ratio between W4/H$\alpha$ luminosities for this
subsample is about ten times larger than what is seen in face-on galaxies,
consistent with larger extinction of the H$\alpha$ emission in edge-on
systems and with incomplete H$\alpha$ coverage. 

\begin{figure}
\epsscale{.90}
\plotone{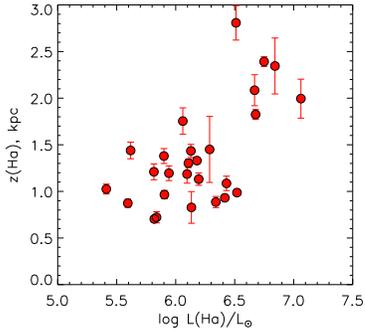}
\caption{The vertical scale height of the ionized gas versus the total $H\alpha$ luminosity in the galaxies.
\label{fig4a}}
\end{figure}

The bottom panel in Figure~\ref{fig4} indicates that there is no
correlation ($cc$ = -0.04) between the rotation velocity lag and the
star formation rate traced by the IR luminosity, a result similar to
the $H\alpha$ data. The lack of the clear correlation between the star
formation rate and the lag amplitude is not necessary an indicator
that there must be an external source of ionization of the gas at high
altitudes. We observe a correlation ($cc$ = 0.55)
between the vertical scale heights of the ionized gas and the star
formation rates, as probed by the total $H\alpha$ luminosity, see
Figure~\ref{fig4a}. This correlation suggests that internal galactic sources
likely play a large role in ionizing gas at high altitudes consistent with
previous studies.

\begin{figure}
\epsscale{.90}
\plotone{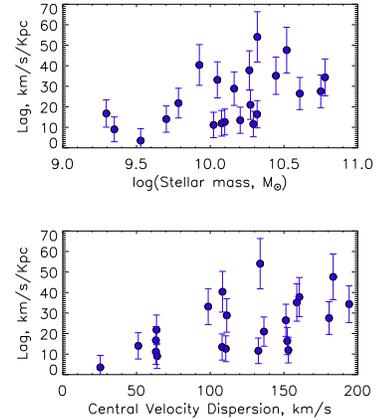} 
\caption{Top panel: the rotation curve lag versus the stellar mass
  estimated in the NSA atlas.  Bottom panel: the rotation curve lag
  versus the the central velocity dispersion in the galaxies .
\label{fig5}}
\end{figure}

Figure~\ref{fig5} demonstrates a noticeable correlation between the
lag and the stellar mass of galaxies ($cc$ = 0.50). The mass in the
upper panel is the stellar mass estimated in the NASA-Sloan Atlas of
galaxies (NSA\footnote{http://nsatlas.org}). The lower panel in
Figure~\ref{fig5} shows the central stellar velocity dispersion from
the NSA ($cc$ = 0.55). Both values are estimated for NSA from earlier SDSS data,
independent of MaNGA. The stellar mass is expected to correlate
strongly with the total dark matter halo mass
(e.g. \citet{behroozi13}).
We also observe significant correlation between
the lag and the maximum rotation velocity in the galaxies $V_{max}$
($cc$ = 0.55, not shown here). The latter is 
expected: large amplitude lags are easier to detect in galaxies with
circular velocities much larger than the total lag.

\begin{figure}
\epsscale{.90}
\plotone{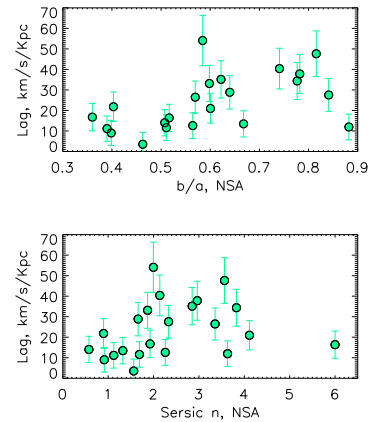} 
\caption{Top panel: the rotation curve lag versus the flatness of the
  galaxy, as estimated by the NSA atlas.  Bottom panel: the rotation
  curve lag versus the galactic Sersic index from the NSA atlas.
\label{fig6}}
\end{figure}

We considered several additional global parameters of the galaxies in
the analysis.  A correlation between the lag value and the axial ratio
of the star light for the galaxies (estimated in NSA) can be seen in
Figure~\ref{fig6} ($cc$ = 0.48).  Less significant is correlation with the
NSA Sersic index of the galaxies ($cc$ = 0.26).  Both parameters are
relevant to morphological type of galaxies, and suggest that we 
tend to observe the largest lag values in early-type disk galaxies.

Active Galactic Nuclei (AGN) could affect our results by contributing
to the $H\alpha$ luminosity via adding a non-star formation-relevant
component, as well as to the motion and ionization status of
extraplanar gas in the central regions.  The former factor can be
neglected: we excluded the central 3 arcsec diameter regions from the
calculation of the $H\alpha$ luminosity and observed the same trends
as in Figure~\ref{fig5}, which is not surprising 
since the mean ratio
of the excluded luminosity to the total is only 0.04 in our galaxies
and an AGN would likely not be directly seen in an edge-on system.

\begin{figure}
\epsscale{.90}
\plotone{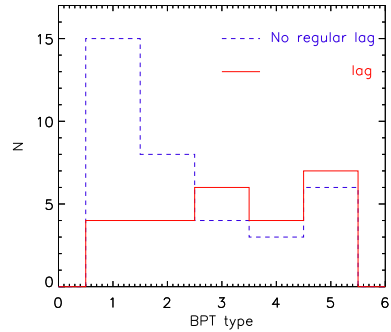}
\caption{The BPT types determined from spectra of the central regions
  in galaxies by the MPA-JHU catalog (see text).  The blue dashed
  curve shows the distribution of galaxies without regular rotation
  lags, while the red solid red line that for for the galaxies with
  regular lags in the extra planar ionized gas.  The BPT types 1-2
  indicate the star forming galaxies, while types 4-5 designate
  AGN/LINER classes (see text for detail).
\label{fig7}}
\end{figure}

Most of our galaxies from both subsamples 
(with and without regular lags) 
were classified with 
"BPT types" \citep[named after Baldwin, Phillips, and Terlevich,][]{BPT} in the MPA-JHU
catalog\footnote{a collaborative project that involved 
Max-Planck Institute for Astronomy and Johns Hopkins University}
 \citep[see e.g.][]{brinchmann04}. Note that the classification
was made with SDSS spectra, which were obtained from the 3-arcsec
central region in galaxies.  The BPT types are: 1- star-forming, 2 -
low signal-to-noise star-forming, 3 - composite, 4 - AGN (excluding
liners), and 5 is a LINER. To summarize them, 1-2 describes the
central star forming dominated galaxies, while 4-5 designate the
presence of a nuclear activity.  Figure~\ref{fig7} demonstrates the
distribution of the galaxies with regular lags and those without
regular lag by BPT type. We see that both galaxy subsamples contain all
BPT types, but the galaxies that do not show a regular lag have a
larger fraction of galaxies dominated by star formation. While this difference
seems suggestive of a role for an AGN, we note that BPT types for
edge-on systems are likely inaccurate due to extinction. The trend
here is likely more consistent with the earlier noted difference in
axial ratios for the two subsamples: the non-regular lag subsample includes
more pure disk galaxies which are less likely to have AGN or LINER
type emission in their central region. It may also be simply related
to the trend with stellar mass: more massive galaxies are more likely
to have an AGN. While the composite and AGN/LINER galaxies all show
regular lags, the star formation-dominated galaxies show mostly no
regular lags. 

Lastly, we considered the distribution of derived lags with the range
in radius over which the fits were executed. We looked at both the
linear range in kpc and the scaled range in radius normalized by
derived radial scale length. For the 25 galaxies, the range in radius
over which the fits were done is between 3 to 20 kpc. This large range
takes place due to various factors: the spread in distance for the galaxies,
their intrinsic disk sizes, and the size of the IFU bundles. This variety
makes it difficult to interpret trends. The one factor that stood out
is that the largest lags (over 30 km s$^{-1}$ kpc$^{-1}$) were found only for
galaxies for which the total fitting radius was less than 10 kpc. It
is possible that this trend is consistent with the general decrease
found in vertical lags with distance from the center for the HI
distribution across single galaxies \citep{zschaechner15a}. That is,
for galaxies that are large enough and well enough covered by the IFU
such that the radial range of our fits extends out to between 10 and
20 kpc on average, we find lower overall lag values than for galaxies
where the fit range is smaller than 10 kpc in radius. The trend,
however, is not seen when the radial range of the fit is expressed in
radius normalized with the disk scalelength. We will need a larger
sample and more deeply consider coverage of the IFU versus total disk
extent of the galaxies to probe such trends further.

\begin{figure}
\epsscale{.90}
\plotone{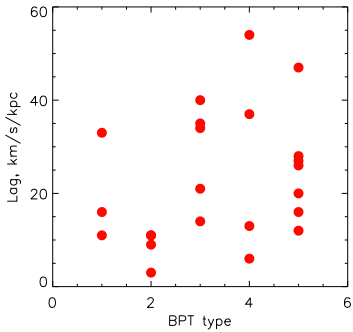}
\caption{The rotation lag versus the BPT type. 
The latter was determined by the MPA-JHU catalog in the central 3-arcsec region.
\label{fig8}}
\end{figure}

The lag amplitude does not correlate strongly with the BPT type as we
show in Figure~\ref{fig8}, although it can be noticed that the
largest lags in our subsample are observed in the galaxies with high BPT
classification, hence potential AGN activity, while small lags (under
20 km s$^{-1}$ kpc$^{-1}$) occur equally often in star-forming and in mild AGN/LINER
galaxies.

\section{Summary and Conclusions}

To summarize, we observe a correlation of the lag
amplitude with the stellar mass (and therefore likely also with halo
mass), with the central velocity dispersion, and also with galaxy
type traced by the stellar disk flatness and global Sersic index.  We
see no correlation between the lag and the star formation
activity in galaxies in the plots of the lag versus
the $H\alpha$ and IR luminosity in Figure~\ref{fig4}. 
According to Figure~\ref{fig6},  late
type disk galaxies tend to have smaller lags, 
and more often than the early-type disk systems
they show no regular lags.  It is plausible that larger lags in early
type disk galaxies with higher mass may be related to the higher
infalling-to-local gas ratio in these systems compared to lower mass
galaxies. Such accreting gas may come from the intergalactic medium
or, alternatively, such massive galaxies are more likely to have
prominent intrinsic hot gaseous halos that may interact with the
disk-halo gas to produce lagging velocities. \citet{marinacci11}
modeled the effect of such a hot halo on the extraplanar gas
kinematics. They found that interaction at the interface between
clouds and the hot halo may lead to cooling of gas from the hot gas
reservoir and subsequent infall into disks. 
\citet{fraternali13} showed that supernova driven galactic 
fountains can transport gas from hot halo down to the galactic disk.
Whether the accretion is
from a pre-existing hot gaseous halo or directly from the IGM, the
lags we observe may provide indirect evidence for accretion of gas.
Since the external gas would have slower rotation with respect to the
gas in galactic disks, accretion of significant amounts of the gas will
create noticeable lags, i.e. the
higher the ratio of infalling-to-local gas, the larger the lag we
observe. 
Although this explanation is plausible, additional
observations and simulations are necessary to confirm.

We performed modeling of the ionized gas radial velocities and
emission line fluxes provided by MaNGA data analysis pipeline MPL-4.
Our models were designed to derive the vertical lag of the ionized gas
rotation velocity that is observed in velocity fields of the gas.

We selected a sample of 67 edge-on galaxies from more than 1400
galaxies observed by MaNGA. In 25 galaxies of this sample (i.e. in
37\%) we observe a regular lag of the rotation curve in the
extraplanar gas.  We derived the lag in the 25 galaxies that span a
wide range of masses and luminosities.

We observe no correlation 
between the amplitude of the lag
and galactic star formation activity traced by $H\alpha$ and IR
luminosity.  At the same time, the lag demonstrates a noticeable
correlation with galactic stellar mass, hence dark matter halo mass,
and central stellar velocity dispersion. The lags are larger in
early-type disk galaxies, whereas smaller lag values are more typical
for late-type galaxies. Early-type galaxies are known to be gas-poor,
while late-type galaxies contain more gas.
We interpret our results as evidence for a
higher ratio of infalling-to-local gas in early-type galaxies and, in
general, the data may point at the importance of gas infall in the
galactic gas balance.

As the MaNGA observing campaign proceeds, we expect to increase our
sample of the galaxies with measured lags by a factor of five (given
MaNGA observing goals), and to build the largest ever sample of
galaxies available for statistical studies of the rotation velocity
lags in the extraplanar ionized gas. 
Accretion from an IGM might not lead to symmetric velocity lags, whereas interaction
with hot halo gas might. 
Our future studies will therefore also in
more detail address the properties of the galaxies that do not show
regular lag and study the possible difference in the lags above and
below disks to distinguish between the different scenarios.

\acknowledgments
This material is based on work partially supported by the National Science 
Foundation under Grant No. AST-1615594 to RAMW.
AD acknowledges support from The Grainger Foundation. 
DB acknowledges support from RSF grant RSCF-14-50-00043. 
We appreciate anonymous referee for valuable suggestions that improved the paper. 

SDSS- IV acknowledges support and resources from the Center for High-Performance 
Computing at the University of Utah. The SDSS web site is www.sdss.org.

SDSS-IV is managed by the Astrophysical Research Consortium for the Participating Institutions of the SDSS Collaboration including the Brazilian Participation Group, the Carnegie Institution for Science, Carnegie Mellon University, the Chilean Participation Group, the French Participation Group, Harvard-Smithsonian Center for Astrophysics, Instituto de Astrof\'isica de Canarias, The Johns Hopkins University, Kavli Institute for the Physics and Mathematics of the Universe (IPMU) / University of Tokyo, Lawrence Berkeley National Laboratory, Leibniz Institut f\"ur Astrophysik Potsdam (AIP), Max-Planck-Institut f\"ur Astronomie (MPIA Heidelberg), Max-Planck-Institut f\"ur Astrophysik (MPA Garching), Max-Planck-Institut f\"ur Extraterrestrische Physik (MPE), National Astronomical Observatory of China, New Mexico State University, New York University, University of Notre Dame, Observatrio Nacional / MCTI, The Ohio State University, Pennsylvania State University, Shanghai Astronomical Observatory, United Kingdom Participation Group, Universidad Nacional Aut\'onoma de M\'exico, University of Arizona, University of Colorado Boulder, University of Oxford, University of Portsmouth, University of Utah, University of Virginia, University of Washington, University of Wisconsin, Vanderbilt University, and Yale University.

{}

\clearpage
\begin{table} 
\begin{center}
\begin{tabular}{lccrrccc}
\tableline\tableline
MANGA-ID & RA(J2000) & Dec(J2000) & Lag & e(Lag) & $z_0(H\alpha)$ & $h(H\alpha)$ & log L(H$\alpha$)/L$_{\odot}$\\
                   &    deg          &     deg         & km s$^{-1}$ kpc$^{-1}$ & km s$^{-1}$ kpc$^{-1}$  &  kpc    &  kpc & dex\\
\tableline
1-339166 & 116.389689 &  45.772328 & 28 &  8 &  1.3 &  2.5 &  6.1 \\
 1-43672 & 116.787163 &  42.078716  & 16 &  7 &  1.0 &  3.0 &  5.9 \\
1-339138$^1$ & 117.318153 &  46.205040 & 11 &  6 &  0.9 &  1.9 &  6.4 \\
 1-44117 & 118.863856 &  44.029330  & 13 &  6 &  1.2 &  4.2 &  5.9 \\
 1-44479 & 119.822532 &  42.008537  & 34 &  9 &  1.5 &  1.5 &  6.3 \\
1-605549 & 134.284434 &  51.471546 &  4  &  6 &  1.0 &  6.1 &  5.4 \\
1-216973 & 135.833415 &  40.464713 & 27 &  8 &  2.8 &  6.4 &  6.5 \\
1-137844 & 139.427012 &  44.100687 & 48 & 11 &  1.4 &  5.4 &  5.6 \\
1-138034 & 144.846118 &  47.126864 & 35 &  9 &  0.9 &  2.8 &  6.3 \\
1-275456 & 159.382717 &  43.653764 & 14 &  6 &  1.3 &  1.6 &  6.2 \\
1-487452 & 166.631358 &  22.777630 & 40 &  10 &  0.8 &  1.2 &  6.1 \\
1-282144 & 184.592510 &  46.155351 & 11 &  6 &  1.8 &  8.8 &  6.7 \\
1-258777 & 186.394835 &  45.431737 & 22 &  7 &  1.1 &  3.5 &  6.4 \\
1-258988 & 187.722485 &  45.219432 & 12 &  6 &  2.1 &  3.4 &  6.7 \\
1-284197 & 194.885392 &  42.758926 & 17 &  7 &  1.2 &  3.3 &  5.8 \\
1-591474 & 197.580687 &  47.124056 & 14 &  6 &  1.2 &  2.5 &  6.1 \\
1-285031 & 198.701372 &  47.351547 & 33 &  9 &  1.0 &  3.0 &  6.5 \\
1-320642 & 214.375198 &  47.714958 &  9 &  6 &  1.1 &  4.8 &  6.2 \\
1-134848 & 244.331994 &  43.479672 & 12 &  6 &  2.3 &  7.3 &  6.8 \\
1-633994 & 247.419951 &  40.686954 & 22 &  7 &  1.4 & 11.0 &  5.9 \\
1-594602 & 249.471810 &  39.249823 & 54 & 12 &  1.8 & 10.0 &  6.1 \\
 1-95111 & 250.668078 &  40.169087  & 29 &  8 &  1.4 &  6.0 &  6.1 \\
1-561039$^1$ & 258.271095 &  35.268616 & 24 &  8 &  0.7 &  1.0 &  5.8 \\
 1-23891 & 260.228215 &  57.097956  & 38 &  10 &  0.9 &  4.5 &  5.6 \\
1-114334$^1$ & 324.259708 &  11.906203  & 73 & 16 &  0.7 &  4.3 &  5.8 \\
\tableline   
\end{tabular}
\caption{The rotation velocity lag and parameters of our MaNGA galaxies subsample: 
MaNGA ID, equatorial coordinates, lag amplitude and its uncertainty, the vertical and radial
scales determined from the $H\alpha$ images, and the total $H\alpha$ luminosity. \\
$^1$ - three galaxies with inclination angles less than 85$^o$ are excluded from the 
figures with the lag in this paper, see text.
\label{tab1} }
\end{center}
\end{table}

\end{document}